\documentclass[twocolumn,prd,superscriptaddress,showpacs,floatfix,%
preprintnumbers,nofootinbib]{revtex4}

\usepackage{epsfig}
\usepackage{bm}

\begin{document}

\title{Self-induced decoherence in dense neutrino gases}

\author{Georg G.~Raffelt}
\affiliation{Max-Planck-Institut f\"ur Physik
(Werner-Heisenberg-Institut), F\"ohringer Ring 6, 80805 M\"unchen,
Germany}

\author{G{\"u}nter Sigl}
\affiliation{APC~\footnote{UMR 7164 (CNRS, Universit\'e Paris 7,
CEA, Observatoire de Paris)} (AstroParticules et Cosmologie),
10, rue Alice Domon et L\'eonie Duquet, 75205 Paris Cedex 13, France}

\date{22 January 2007}

\preprint{MPP-2007-6}

\begin{abstract}
Dense neutrino gases exhibit collective oscillations where
``self-maintained coherence'' is a characteristic feature, i.e.,
neutrinos of different energies oscillate with the same frequency.
In a non-isotropic gas, however, the flux term of the
neutrino-neutrino interaction has the opposite effect of causing
kinematical decoherence of neutrinos propagating in different
directions, an effect that is at the origin of the ``multi-angle
behavior'' of neutrinos streaming off a supernova core. We cast the
equations of motion in a form where the role of the flux term is
manifest. We study in detail the symmetric case of equal neutrino
and antineutrino densities where the evolution consists of
collective pair conversions (``bipolar oscillations''). A gas of
this sort is unstable in that an infinitesimal anisotropy is enough
to trigger a run-away towards flavor equipartition. The
``self-maintained coherence'' of a perfectly isotropic gas gives way
to ``self-induced decoherence.''
\end{abstract}


\maketitle

\section{Introduction}                        \label{sec:introduction}

Neutrino oscillations between two flavors in vacuum are governed by
the frequency
\begin{equation}
\omega=\frac{\Delta m^2}{2E}
\end{equation}
where $E$ is the energy of a given mode. Therefore, if the energy
spectrum is broad, the energy dependence of the oscillation
frequency quickly leads to kinematical decoherence, i.e., along a
neutrino beam the overall flavor content quickly approaches an
average value.

The situation changes radically when neutrinos themselves provide a
significant refractive effect, leading to collective oscillation
modes~\cite{Pantaleone:1992eq, Samuel:1993uw, Kostelecky:1993yt,
Kostelecky:1993dm, Kostelecky:1994ys, Kostelecky:1995dt,
Kostelecky:1995xc, Samuel:1996ri, Kostelecky:1996bs,
Pantaleone:1998xi, Sawyer:2005jk, Pastor:2001iu} that can be of
practical interest in the early universe~\cite{Lunardini:2000fy,
Dolgov:2002ab, Wong:2002fa, Abazajian:2002qx} or in core-collapse
supernovae~\cite{Pantaleone:1994ns, Qian:1994wh, Sigl:1994hc,
Pastor:2002we, Balantekin:2004ug, Fuller:2005ae, Duan:2005cp,
Duan:2006an, Duan:2006jv, Hannestad:2006nj}. Defining the parameter
\begin{equation}
\mu=\sqrt{2}\,G_{\rm F} n_\nu\,,
\end{equation}
the neutrino gas is ``dense'' when $\omega\alt\mu$, i.e., when a
typical neutrino-neutrino interaction energy exceeds the energy
corresponding to the vacuum oscillation frequency. When this
condition is satisfied, collective effects are important, even if
the ordinary matter effect is much larger than that from the
neutrino-neutrino interaction~\cite{Duan:2005cp, Hannestad:2006nj}.
One characteristic feature of collective oscillations is the
phenomenon of ``self-maintained
coherence''~\cite{Kostelecky:1995dt}, meaning that all modes
oscillate with the same frequency even though the energy spectrum
may be broad.

It was recently stressed, however, that this is not the complete
story~\cite{Sawyer:2005jk, Duan:2006an}. Perhaps the most
interesting case for collective effects is provided by neutrinos
streaming off a supernova core, a situation that is far from
isotropic. The current-current nature of the weak-interaction
Hamiltonian implies that the interaction energy between particles of
momenta ${\bf p}$ and ${\bf q}$ is proportional to $(1- {\bf v}_{\bf
p}\cdot{\bf v}_{\bf q})$ where ${\bf v}_{\bf p}={\bf p}/E$ is the
velocity. If the medium is isotropic, the ${\bf v}_{\bf p}\cdot{\bf
v}_{\bf q}$ term averages to zero, but if there is a net current,
test particles moving in different directions experience different
refractive effects. Therefore, neutrinos moving in a background with
a net current will decohere between different directions of motion,
even if the energy spectrum is monochromatic.

To avoid confusion about terminology we stress that we always mean
``kinematical decoherence'' when we say ``decoherence.'' Different
modes oscillate differently, leading to ``de-phasing'' and thus to
the depolarization of the overall ensemble. If we use the common
language of polarization vectors ${\bf P}_{\bf p}$ for each mode
${\bf p}$, then the length of each ${\bf P}_{\bf p}$ is conserved,
whereas the length of the overall polarization vector ${\bf
P}=\sum{\bf P}_{\bf p}$ shrinks (kinematical decoherence) or is
conserved (kinematical coherence). The effect of {\it dynamical\/}
decoherence, caused by collisions among the neutrinos or with a
thermal background medium, is that each individual polarization
vector ${\bf P}_{\bf p}$ shrinks, i.e., neutrinos in individual
modes cannot be represented as pure states. This effect, relevant
for open quantum systems, does not occur in our case where
oscillations are the only form of evolution.

The multi-angle decoherence effect becomes nontrivial in the most
interesting case when the ``background current'' consists of the
neutrinos themselves as for neutrinos streaming off a supernova
core. Numerical examples reveal significant decoherence effects, but
on the other hand they also show collective modes very similar to
the isotropic (``single-angle'') case~\cite{Duan:2006an}. The
overall behavior is determined by a complicated interplay between
the collective evolution and decoherence.

Rather than trying to understand the supernova case in its full
complexity, we here take the opposite approach and study the
simplest example that shows decoherence caused by the
neutrino-neutrino multi-angle effect. Therefore, we consider a dense
neutrino gas that is monochromatic, symmetric (equal $\nu$ and
$\bar\nu$ densities), and homogeneous, but not isotropic. Even this
simple model has a rich phenomenology that helps one to develop a
better understanding of the full problem. On the other hand, it is
simple enough that the most puzzling aspects of its behavior are
analytically tractable.

A symmetric $\nu\bar\nu$ gas oscillates in the ``bipolar mode''
where pairs of neutrinos of a given flavor coherently oscillate into
the other flavor and back with the ``bipolar
frequency''~\cite{Kostelecky:1995dt, Samuel:1996ri, Duan:2005cp,
Hannestad:2006nj}
\begin{equation}\label{eq:kappa}
\kappa=\sqrt{2\omega\mu}\,.
\end{equation}
As an example we consider a dense neutrino gas initially consisting
of $\nu_e\bar\nu_e$ and assume that the mixing angle $\Theta$ with
another flavor is small, a situation that could represent
13-oscillations. We characterize the overall flavor content of the
ensemble in the usual way by a polarization vector where the
positive $z$-direction corresponds to the $\nu_e$-flavor whereas the
mass direction is defined by a unit vector that we call ${\bf B}$.
It is tilted against the $z$-direction by $2\Theta$ and we always
choose it to lie in the $x$-$z$-plane with $B_z<0$ for the normal
hierarchy and $B_z>0$ for the inverted hierarchy.

\begin{figure}
\begin{center}
\epsfig{file=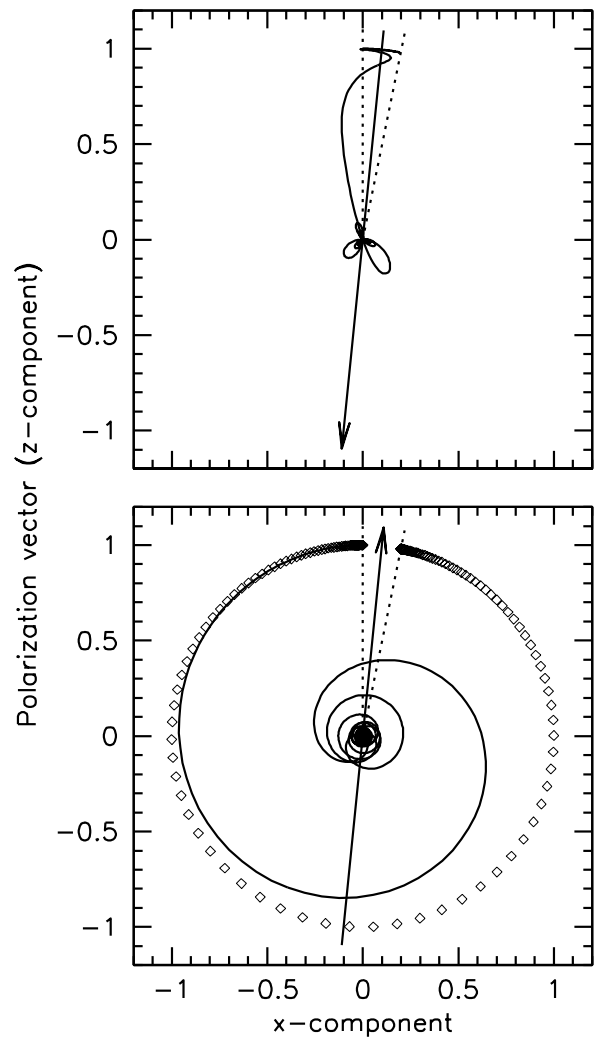,width=0.9\columnwidth}
\end{center}
\caption{Trajectory of the overall polarization vector in the
``half-isotropic case'' for the normal hierarchy (top) and inverted
hierarchy (bottom) with $\sin2\Theta=0.1$. The arrow is the ${\bf
B}$ direction. In the isotropic case (no decoherence), the dotted
lines represent the largest pendulum excursion of $2\Theta$ on
either side of ${\bf B}$. The diamonds in the lower panel represent
the pendular motion in time intervals of
$0.01\,(2\pi/\kappa)$.}\label{fig:trajectory}
\end{figure}

In vacuum, the polarization vectors precess around ${\bf B}$ with
frequency $\omega$. In a dense neutrino gas with $\Theta\ll1$, their
motion is largely confined to the $x$-$z$-plane where they perform
pendular motions with frequency $\kappa$ and where $-{\bf B}$ is the
``force direction.'' The maximum excursion is $2\Theta$, indicated
by the dotted lines in Fig.~\ref{fig:trajectory} where we show the
trajectory in the $x$-$z$-plane. For the normal hierarchy, the
pendulum swings between the dotted lines. For the inverted hierarchy
(bottom), it performs almost full-circle oscillations as indicated
by the diamonds.

If the gas is not isotropic, the picture changes completely. Taking
the ``half-isotropic'' case as a generic example where only one
half-space of momentum modes is occupied (neutrinos streaming off a
plane surface), we show in Fig.~\ref{fig:trajectory} the trajectory
of the average polarization vector. For both hierarchies, its length
shrinks to zero so that the final state of the ensemble is an equal
mixture of both flavors. In other words, the ensemble quickly
decoheres, independently of the mass hierarchy. This behavior is
perhaps intuitive because we have assumed a large neutrino current.

However, this effect is self-accelerating in that it is triggered
even by the smallest nonzero anisotropy. Whereas for perfect
isotropy the neutrino-neutrino interaction provides for
self-maintained coherence between modes of different energy, even an
infinitesimal anisotropy has the opposite effect of causing
self-induced decoherence between modes with different propagation
directions. The average flavor content of the ensemble always
approaches an equal mixture of both flavors. This instability is a
nonlinear effect caused by the neutrino-neutrino interaction. In
other words, the pendular motion of the isotropic ensemble is an
unstable limit cycle, whereas the equal flavor mixture is a stable
fixed point.

Our goal is to explain this behavior. As a first step we formulate
in Sect.~\ref{sec:eom} the equations of motion for the multi-angle
system in terms of the multipole components of the neutrino angular
distribution. The special role of the flux (the dipole term of the
angular distribution) becomes manifest. We begin with the full
$N$-flavor problem and then consider two flavors in terms of
polarization vectors. In Sect.~\ref{sec:anisotropy} we study the
role of a large anisotropy in causing decoherence. Our main result
is developed in Sect.~\ref{sec:smallanisotropy} where we show that a
$\nu\bar\nu$ gas is unstable in flavor space and always decoheres if
there is an infinitesimally small initial anisotropy. We conclude in
Sec.~\ref{sec:conclusions} with a summary and outlook.

\section{Multi-angle equations of motion}              \label{sec:eom}

\subsection{General equations of motion}

A statistical ensemble of unmixed neutrinos is represented by the
occupation numbers $f_{\bf p}=\langle a^\dagger_{\bf p} a_{\bf
p}\rangle$ for each momentum mode ${\bf p}$, where $a^\dagger_{\bf
p}$ and $a_{\bf p}$ are the relevant creation and annihilation
operators and $\langle\ldots\rangle$ is the expectation value. The
corresponding expression for antineutrinos is ${\bar f}_{\bf
p}=\langle {\bar a}^\dagger_{\bf p} {\bar a}_{\bf p}\rangle$, where
here and henceforth overbarred quantities refer to antiparticles.

For several flavors, the occupation numbers are generalized to
``matrices of densities'' in flavor space~\cite{Dolgov:1980cq,
Rudzsky:1990, Sigl:1992fn, mckellar&thomson}
\begin{equation}\label{eq:densitymatrixdefinition}
 (\varrho_{\bf p})_{ij}=
 \langle a^\dagger_{j} a_{i}\rangle_{\bf p}
 \hbox{\quad and\quad}
 (\bar \varrho_{\bf p})_{ij}=
 \langle \bar a^\dagger_{i}\,\bar a_{j}\rangle_{\bf p}
\end{equation}
that really are matrices of occupation numbers. The reversed order
of the flavor indices $i$ and $j$ in the r.h.s.\ for antineutrinos
ensures that under a flavor transformation, $\varrho_{\bf p}\to
U\varrho_{\bf p}U^\dagger$, antineutrinos transform in the same way,
$\bar\varrho_{\bf p}\to U\bar\varrho_{\bf p}U^\dagger$. Sums and
differences of $\varrho_{\bf p}$ and $\bar\varrho_{\bf p}$ then
transform consistently. For the seemingly intuitive equal order of
flavor indices that is almost universally used in the literature,
$\varrho_{\bf p}$, $\bar\varrho_{\bf p}$, $\varrho^*_{\bf p}$ and
$\bar\varrho^*_{\bf p}$ all appear in the equations instead of
$\varrho_{\bf p}$ and $\bar\varrho_{\bf p}$ alone.

Analogous matrices $(\ell_{\bf p})_{ij}$ can be defined for the
charged leptons.

We assume that the neutrino ensemble is completely described by these
one-particle occupation number matrices, i.e., that genuine many-body
effects play no role~\cite{Friedland:2003dv, Friedland:2003eh,
Friedland:2006ke}. In this case flavor oscillations of an ensemble of
neutrinos and antineutrinos are described by an equation of
motion for each mode~\cite{Dolgov:1980cq, Rudzsky:1990, Sigl:1992fn,
mckellar&thomson}
\begin{widetext}
\begin{eqnarray}\label{eq:eom1}
 {\rm i}\,\partial_t \varrho_{\bf p}&=&
 +\left[\Omega_{\bf p},\varrho_{\bf p}\right]
 +\sqrt{2}\,G_{\rm F}\left[
 \int\!\frac{d^3{\bf q}}{(2\pi)^3}
 \left(\ell_{\bf q}-\bar\ell_{\bf q}
 +\varrho_{\bf q}-\bar\varrho_{\bf q}\right)
 (1-{\bf v}_{\bf q}\cdot{\bf v}_{\bf p})
 ,\varrho_{\bf p}\right],
 \nonumber\\
 {\rm i}\,\partial_t\bar\varrho_{\bf p}&=&
 -\left[\Omega_{\bf p},\bar\varrho_{\bf p}\right]
 +\sqrt{2}\,G_{\rm F}\left[
 \int\!\frac{d^3{\bf q}}{(2\pi)^3}
 \left(\ell_{\bf q}-\bar\ell_{\bf q}
 +\varrho_{\bf q}-\bar\varrho_{\bf q}\right)
 (1-{\bf v}_{\bf q}\cdot{\bf v}_{\bf p})
 ,\bar\varrho_{\bf p}\right],
\end{eqnarray}
where $[{\cdot},{\cdot}]$ is a commutator and $G_{\rm F}$ is the
Fermi constant. For ultrarelativistic neutrinos, the matrix of
vacuum oscillation frequencies, expressed in the mass basis, is
$\Omega_{\bf p}={\rm diag}(m_1^2,m_2^2,m_3^2)/2p$ with $p=|{\bf
p}|$. Further, ${\bf v}_{\bf p}={\bf p}/E_{\bf p}$ is the velocity
of a particle (neutrino or charged lepton) with momentum ${\bf p}$.

The total conserved energy of the neutrino ensemble is an important
quantity for understanding its evolution~\cite{Duan:2005cp,
Hannestad:2006nj}. We find for the energy density
\begin{eqnarray}\label{eq:energydensity}
 \varepsilon&=&
 {\rm Tr}\biggl\lbrace\int\!\frac{d^3{\bf p}}{(2\pi)^3}\,
 \Omega_{\bf p}(\varrho_{\bf p}+\bar\varrho_{\bf p})
 +\sqrt2\,G_{\rm F}
 \int\!\frac{d^3{\bf p}}{(2\pi)^3}
 \int\!\frac{d^3{\bf q}}{(2\pi)^3}\,
 (\ell_{\bf q}-\bar\ell_{\bf q})
 (\varrho_{\bf p}-\bar\varrho_{\bf p})
 (1-{\bf v}_{\bf q}\cdot{\bf v}_{\bf p})
 \nonumber\\
 &&\kern7em{}+\frac{\sqrt2\,G_{\rm F}}{2}
 \left[\int\!\frac{d^3{\bf p}}{(2\pi)^3}\,
 (\varrho_{\bf p}-\bar\varrho_{\bf p})\right]^2
 -\frac{\sqrt2\,G_{\rm F}}{2}\left[
 \int\!\frac{d^3{\bf p}}{(2\pi)^3}\,
 (\varrho_{\bf p}-\bar\varrho_{\bf p}){\bf v}_{\bf p}\right]^2
 \biggr\rbrace\,.
\end{eqnarray}
\end{widetext}
This quantity actually represents the energy {\it shift\/} caused by
the neutrino masses and by neutrino interactions. It is
straightfoward to show that indeed $\dot\varepsilon=0$ by taking the
time derivative on the r.h.s.\ of Eq.~(\ref{eq:energydensity}),
inserting the equations of motion Eq.~(\ref{eq:eom1}), and using
cyclic permutations of matrices under the trace.

For completeness we mention that one can also define the entropy
density~\cite{Sigl:1992fn}
\begin{eqnarray}
 s&=&-
 \int\frac{d^3{\bf p}}{(2\pi)^3}\,
 {\rm Tr}\bigl[
 \varrho_{\bf p}\ln(\varrho_{\bf p})
 +(1-\varrho_{\bf p})\ln(1-\varrho_{\bf p})
 \nonumber\\
 &&\kern3em{}
 +\bar\varrho_{\bf p}\ln(\bar\varrho_{\bf p})
 +(1-\bar\varrho_{\bf p})\ln(1-\bar\varrho_{\bf p})
 \bigr]\,.
\end{eqnarray}
This expression is well defined because $\varrho_{\bf p}$ and
$1-\varrho_{\bf p}$ are positive semi-definite matrices. In our case
where the equation of motion for each $\varrho_{\bf p}$ is of the
form ${\rm i}\partial_t\varrho_{\bf p}=[H,\varrho_{\bf p}]$, the
entropy density is conserved. Oscillations alone do not lead to a
loss of information. Kinematical decoherence does not lead to an
increase of entropy.

\subsection{Axial symmetry}

Since we wish to study the simplest example that shows nontrivial
multi-angle effects, we now restrict the neutrino energy
distribution to be monochromatic and the geometry to axial symmetry.
We then integrate the matrices of densities over all variables
except $u=\cos\theta_{\bf p}$ where the angle is relative to the
direction of symmetry. Therefore, we consider the matrices
\begin{equation}
 \varrho_u=\frac{1}{n_{\nu}}
 \int\frac{dp\,p^2\,d\varphi}{(2\pi)^3}\,\varrho_{\bf p}\,,
\end{equation}
implying the normalization
\begin{equation}
 \varrho_0\equiv\frac{1}{n_{\nu}}
     \int\frac{d^3{\bf p}}{(2\pi)^3}\,\varrho_{\bf p}
     =\int_{-1}^{+1}du\,\varrho_u\,.
\end{equation}
For convenience we have arbitrarily normalized the $\varrho_u$
matrices to the neutrino density.

We further define the matrix representing the particle flux along
the direction of symmetry,
\begin{equation}
 \varrho_1\equiv\int_{-1}^{+1}du\,\varrho_u\,v_u\,,
\end{equation}
where $v_u$ is the velocity along the symmetry direction. For
relativistic particles $v_u=\cos\theta_u=u$.

For charged leptons we proceed in the same way, except that we
normalize to the electron density $n_e$.

We denote with $\Omega$ the matrix of vacuum oscillation frequencies
for our fixed neutrino energy $E$. The equations of motion then
simplify to
\begin{eqnarray}\label{eq:eom2}
 {\rm i}\,\dot\varrho_{u}&=&
 \left[\Omega,\varrho_{u}\right]
 +\left[\lambda\left(\ell-\bar\ell)_0
 +\mu(\varrho-\bar\varrho\right)_0,\varrho_u\right]
 \nonumber\\
 &&\kern2.8em{}-\left[\lambda\left(\ell-\bar\ell)_1
 +\mu(\varrho-\bar\varrho\right)_1,u\varrho_u\right]\,,
\end{eqnarray}
where $\lambda=\sqrt2\,G_{\rm F}n_{e}$. The same equation applies
for $\bar\rho_u$ except for a sign change of the vacuum oscillation
term.

Evidently the isotropic part of the medium (index~0) affects all
modes in the same way and is ultimately responsible for the
phenomenon of self-maintained coherence. On the other hand, the flux
term (index~1) involves a factor $u=\cos\theta$ for every mode $u$.
Even in the absence of a neutrino-neutrino term, a charged-lepton
flux alone causes a trivial multi-angle decoherence effect.

The conserved energy of the axially symmetric neutrino ensemble is
\begin{eqnarray}
 E&=&{\rm Tr}\left[\Omega(\varrho_0+\bar\varrho_0)\right]
 \nonumber\\
 &&{}+\lambda\,{\rm Tr}\left[
 (\ell-\bar\ell)_0(\varrho-\bar\varrho)_0
 -(\ell-\bar\ell)_1(\varrho-\bar\varrho)_1\right]
 \nonumber\\
 &&{}+\frac{\mu}{2}\,{\rm Tr}\left[
 (\varrho-\bar\varrho)_0^2
 -(\varrho-\bar\varrho)_1^2\right]\,.
\end{eqnarray}
Because of our normalization of the $\varrho_u$ matrices the
quantity $E=\varepsilon/n_{\nu}$ is the energy per $\nu$.

\subsection{Expansion in Legendre polynomials}

The structure of these equations becomes more transparent if we
expand the angular dependence of the matrices in Legendre
polynomials. (Had we not assumed axial symmetry, spherical harmonics
would be the appropriate basis.) The first few Legendre
polynomials~are
\begin{eqnarray}
L_0(u)&=&1\,,\nonumber\\
L_1(u)&=&u\,,\nonumber\\
L_2(u)&=&{\textstyle\frac{1}{2}}(3u^2-1)\,.
\end{eqnarray}
We thus define
\begin{equation}\label{eq:expansion1}
 \varrho_n=\int_{-1}^{+1}du\,\varrho_u L_n(u)\,.
\end{equation}
The normalisation
\begin{equation}
 \int_{-1}^{+1}du\,L_m(u)L_n(u)
 =\frac{2}{2n+1}\,\delta_{mn}
\end{equation}
implies that the original function is
\begin{equation}\label{eq:expansion2}
 \varrho_u=\sum_{n=0}^{\infty}
 (n+\textstyle{\frac{1}{2}})\varrho_n L_n(u)\,.
\end{equation}
The previously defined overall density $\varrho_0$ and the flux
$\varrho_1$ are but the zeroth and first case of
Eq.~(\ref{eq:expansion1}).

Multiplying both sides of Eq.~(\ref{eq:eom2}) with $L_n(u)$ and
integrating over $du$ leads to the equation of motion
\begin{eqnarray}
 {\rm i}\,\dot\varrho_n&=&
 \left[\Omega,\varrho_n\right]
 +\mu\left[(\varrho-\bar\varrho)_0,\varrho_n\right]
 \nonumber\\
 &&{}-\mu\left[(\varrho-\bar\varrho)_1,
 {\textstyle\int_{-1}^{+1}}du\,u\varrho_u L_n(u)\right]\,.
\end{eqnarray}
Here and henceforth we no longer show the ordinary matter term. Its
structure is similar to the neutrino-neutrino term so that it is
easily reinstated.

With the expansion Eq.~(\ref{eq:expansion2}) the remaining integral
is
\begin{eqnarray}
 \int_{-1}^{+1}\!du\,u\,\varrho_u L_n(u)&&\nonumber\\
 &&\kern-8em{}=\sum_{m=0}^{\infty} (m+{\textstyle\frac{1}{2}})
 \varrho_m \int_{-1}^{+1}\!du\,u\,L_m(u)L_n(u)\,.
\end{eqnarray}
We use the  ``dipole matrix element'' of the Legendre polynomials
\begin{eqnarray}
 \int_{-1}^{+1}du\,u\,L_m(u)\,L_n(u)&=&
 \frac{2(m+1)}{(2m+1)(2m+3)}\,\delta_{m+1,n}
 \nonumber\\
 &+&\frac{2m}{(2m-1)(2m+1)}\,\delta_{m-1,n}\,.
 \nonumber\\
\end{eqnarray}
With this result the equations of motion are
\begin{eqnarray}\label{eq:eom3}
 {\rm i}\,\dot\varrho_n&=&
 \left[\Omega,\varrho_n\right]
 +\mu\left[(\varrho-\bar\varrho)_0,\varrho_n\right]
 \nonumber\\
 &&{}-\frac{\mu}{2}
 \left[(\varrho-\bar\varrho)_1,
 (a_n\varrho_{n-1}+b_n\varrho_{n+1})\right]\,,
\end{eqnarray}
where
\begin{eqnarray}\label{eq:abcoefficients}
 a_n&=&\frac{2n}{2n+1}\kern0.7em
 =1-\frac{1}{2n+1}\,,\nonumber\\
 b_n&=&\frac{2(n+1)}{2n+1}
 =1+\frac{1}{2n+1} \,.
\end{eqnarray}
Note that the equation for $\varrho_0$ is consistent even though the
quantity $\varrho_{-1}$ is not defined because the coefficient
$a_n=n/(2n+1)=0$ for $n=0$.

The equations of motion Eq.~(\ref{eq:eom3}) together with the
corresponding equations for $\bar\varrho_n$ (sign change for the
vacuum term) form a closed set of equations. If we also include the
ordinary matter term they are equivalent to Eq.~(\ref{eq:eom2}) for
the momentum-space matrices $\varrho_u$.

One important difference is that ${\rm Tr}(\varrho^2_u)$ is
conserved, reflecting the absence of decoherence for individual
momentum modes, whereas ${\rm Tr}(\varrho^2_n)$ is not in general
conserved, reflecting the effect of kinematical decoherence. We note
that in general $\partial_t{\rm Tr}({\sf A}^2)=0$ when the equation
of motion is of the form ${\rm i}\partial_t {\sf A}=[{\sf A},{\sf
H}]$. This is the case for $\varrho_u$, but not for the multipole
matrices $\varrho_n$.

Of course, in a numerical implementation the series $\varrho_n$ has
to be truncated at some value $n_{\rm max}$, leading to limited
angular resolution. This is analogous to the coarse graining of
phase space required for the $\rho_u$ where one needs to use
discrete angular bins of nonzero width.

The equations of motion for the flux terms are special in that they
involve one power of $\varrho_1$ or $\bar\varrho_1$ in each term of
the equation. Therefore, if initially there is no flux term
($\varrho_1=\bar\varrho_1=0$), none will develop. In this case the
equations for $\varrho_0$ and $\bar\varrho_0$ form a closed set,
describing the dynamics of the ``flavor pendulum'' studied in
Ref.~\cite{Hannestad:2006nj}. In addition, the higher multipoles
$\varrho_n$ with $n\geq2$, if initially nonzero, will simply
oscillate under the action of the vacuum term and of the density
term $(\varrho-\bar\varrho)_0$.

\subsection{Diffusion equation in multipole space}

The interpretation of the equations of motion in multipole space
becomes clearer if we observe that for some function $f(x)$ one has
to second order
\begin{eqnarray}
 f(x+\Delta x)=f(x)+f^\prime(x)\Delta x
 +{\textstyle\frac{1}{2}}
 \,f^{\prime\prime}(x)\Delta x^2\,,\nonumber\\
 f(x-\Delta x)=f(x)-f^\prime(x)\Delta x
 +{\textstyle\frac{1}{2}}
 \,f^{\prime\prime}(x)\Delta x^2\,.
\end{eqnarray}
Taking the sum and difference of these equations provides
\begin{eqnarray}
 f(x+\Delta x)+f(x-\Delta x)&=&2f(x)+f^{\prime\prime}(x)\Delta
 x^2\,,
 \nonumber\\
 f(x+\Delta x)-f(x-\Delta x)&=&2f^\prime(x)\Delta x\,.
\end{eqnarray}
With Eq.~(\ref{eq:abcoefficients}) we may write
\begin{equation}\label{eq:diffusion}
 a_n\varrho_{n-1}+b_n\varrho_{n+1}=
 2\varrho_n+
 \frac{2}{2n+1}\,\varrho^\prime_n\Delta n
 +\varrho_n^{\prime\prime}\Delta n^2\,,
\end{equation}
where we interpret a prime as a derivative with respect to $n$ that
is now viewed as a continuous variable. If we further observe that
in the discrete equation $\Delta n=1$ we may write
Eq.~(\ref{eq:eom3}) in the continuous form
\begin{eqnarray}\label{eq:eom4}
 {\rm i}\,\dot\varrho_n&=&
 \left[\Omega,\varrho_n\right]
 +\mu\left[(\varrho-\bar\varrho)_0-(\varrho-\bar\varrho)_1,
 \varrho_n\right]
 \nonumber\\
 &&\kern2.7em{}-\mu
 \left[(\varrho-\bar\varrho)_1,\left(
 \frac{\varrho^\prime_n}{2n+1}+\frac{\varrho^{\prime\prime}_n}{2}
 \right)\right]\,.
\end{eqnarray}
Therefore, the lowest multipole (monopole) is equivalent to the
pendulum in flavor space~\cite{Hannestad:2006nj}, the high
multipoles are akin to a continuous medium carrying flavor waves.
Equation~(\ref{eq:eom4}) is a diffusion equation. The periodic
excitation caused by the flavor pendulum then diffuses to higher
multipoles (smaller scales). The flavor waves as a function of
neutrino direction visible in numerical
simulations~\cite{Duan:2006jv} are the result of this process.

\eject

\subsection{Particle vs.\ lepton number}

In the equations of motion the difference of the particle and
antiparticle densities, $\varrho_{\bf p}-\bar\varrho_{\bf p}$, plays
a special role in that it is this matrix of net lepton number
densities that plays the role of a potential for the other neutrino
modes. As noted by several authors, it is sometimes useful to write
the equations of motion in terms of the sum and difference of
$\varrho_{\bf p}$ and $\bar\varrho_{\bf p}$ instead of these
matrices themselves. We therefore define the matrix of particle
densities (sum~$\sf S$) and the matrix of lepton number densities
(difference~$\sf D$) by virtue of
\begin{eqnarray}
{\sf S}_{\bf p}&=&\varrho_{\bf p}+\bar\varrho_{\bf p}\,,
\nonumber\\
{\sf D}_{\bf p}&=&\varrho_{\bf p}-\bar\varrho_{\bf p}\,.
\end{eqnarray}
Analogous definitions apply to the case of axial symmetry, ${\sf
S}_u$ and ${\sf D}_u$, and their multipole expansion ${\sf S}_n$
and~${\sf D}_n$.

The equations of motion for the particle and lepton number matrices
are found by adding and subtracting the two lines of
Eq.~(\ref{eq:eom1}) and the equivalent equations for axial symmetry
or the multipoles. We find explicitly for the multipoles
\begin{eqnarray}\label{eq:eom5}
{\rm i}\dot{\sf S}_n&=&[\Omega,{\sf D}_n]
+\mu[{\sf D}_0,{\sf S}_n]
-\mu\left[{\sf D}_1,
\frac{a_n{\sf S}_{n-1}+b_n{\sf S}_{n+1}}{2}\right],
\nonumber\\
{\rm i}\dot{\sf D}_n&=&[\Omega,{\sf S}_n]
+\mu[{\sf D}_0,{\sf D}_n]
-\mu\left[{\sf D}_1,
\frac{a_n{\sf D}_{n-1}\!+\!b_n{\sf D}_{n+1}}{2}\right].
\nonumber\\
\end{eqnarray}
One can easily restore the ordinary matter term because it has the
same structure as the neutrino-neutrino term with the substitution
$\mu\to\lambda$ and ${\sf D}_{0,1}\to {\sf L}_{0,1}$, the latter
being the charged-lepton matrix of net lepton-number densities and
the corresponding flux, respectively.

The commutator structure of the r.h.s.\ of Eq.~(\ref{eq:eom5})
implies that its trace vanishes. Therefore, the particle and lepton
numbers for all multipoles, ${\rm Tr}({\sf S}_n)$ and ${\rm Tr}({\sf
D}_n)$, are conserved.

With the expressions Eq.~(\ref{eq:abcoefficients}) for $a_n$ and
$b_n$ the lowest-order multipole equation is
\begin{eqnarray}
{\rm i}\dot{\sf S}_0&=&[\Omega,{\sf D}_0] +\mu[{\sf D}_0,{\sf S}_0]
-\mu[{\sf D}_1,{\sf S}_{1}]\,,
\nonumber\\
{\rm i}\dot{\sf D}_0&=&[\Omega,{\sf S}_0]\,.
\end{eqnarray}
One immediate consequence of the second equation is
\begin{equation}
\partial_t{\rm Tr}(\Omega{\sf D}_0)=0\,,
\end{equation}
i.e., in the mass basis a certain combination of flavor-lepton
numbers is conserved.

The explicit equation for the flux term is
\begin{eqnarray}
 {\rm i}\dot{\sf S}_1&=&[\Omega,{\sf D}_1] +\mu[{\sf D}_0,{\sf S}_1]
 +\frac{\mu}{3}\,[{\sf S}_{0}+2{\sf S}_{2},{\sf D}_1]\,,
 \nonumber\\
 {\rm i}\dot{\sf D}_1&=&[\Omega,{\sf S}_1]
 +\frac{2\mu}{3}\,[{\sf D}_0-{\sf D}_2,{\sf D}_{1}]\,.
\end{eqnarray}
As stated earlier, these equations are special in that they involve
one power of ${\sf S}_1$ or ${\sf D}_1$ in each term. Therefore,
${\sf S}_1$ and ${\sf D}_1$ vanish at all times if they vanish
initially.

In this case, the monopole terms (overall densities) form a closed
set of equations. Defining
\begin{eqnarray}
 {\sf Q}&=&{\sf S}_0-\frac{\Omega}{\mu}\,,
 \nonumber\\
 {\sf D}&=&{\sf D}_0\,,
\end{eqnarray}
the equations of motion take on the simple form
\begin{eqnarray}
 {\rm i}\dot{\sf Q}&=&\mu[{\sf D},{\sf Q}]\,,
 \nonumber\\
 {\rm i}\dot{\sf D}&=&[\Omega,{\sf Q}]\,.
\end{eqnarray}
Conserved quantities are ${\rm Tr}({\sf Q}^2)$, ${\rm Tr}({\sf
DQ})$, ${\rm Tr}({\sf D\Omega})$, and ${\rm Tr}(\Omega{\sf
Q}+\mu{\sf D}^2/2)$. Up to a constant, the last quantity is the
energy per neutrino. For two flavors, these simplifications are at
the origin of the pendulum analogy~\cite{Hannestad:2006nj} where
${\rm Tr}({\sf Q}^2)$ corresponds to its length, ${\rm Tr}({\sf
DQ})$ its spin, and ${\rm Tr}({\sf D\Omega})$ to the orbital angular
momentum around the force direction (the mass direction in flavor
space).

Returning to the general case with arbitrary initial conditions, the
energy per neutrino is
\begin{equation}\label{eq:energy2}
 E={\rm Tr}\left[\Omega{\sf S}_0 +
 \frac{\mu}{2}\left({\sf D}_0^2-{\sf D}_1^2\right)\right]\,.
\end{equation}
Energy conservation $\partial_t E=0$ is easily verified by using the
equations of motion for ${\sf S}_0$, ${\sf D}_0$ and ${\sf D}_1$.

\subsection{Two-flavor case}

In the two-flavor case where the $\varrho_{\bf p}$ are Hermitian
$2\times2$ matrices, a representation in terms of polarization
vectors is useful. A Hermitian $2\times2$ matrix ${\sf A}$ is
represented as
\begin{equation}
 {\sf A}=\frac{{\rm Tr}({\sf A})
 +{\bf A}\cdot\bm{\sigma}}{2}\,,
\end{equation}
where $\bm\sigma$ is the vector of Pauli matrices and ${\bf A}$ the
polarization vector.

The commutation relations of the Pauli matrices imply that an
equation of motion of the form
\begin{equation}
{\rm i}\partial_t{\sf A}=[{\sf B},{\sf C}]
\end{equation}
is represented by
\begin{equation}
 \partial_t{\rm Tr}({\sf A})=0
 \hbox{\quad and\quad}
 \partial_t{\bf A}={\bf B}\times{\bf C}\,.
\end{equation}
Moreover,
\begin{equation}\label{eq:doubletrace}
 {\rm Tr}({\sf AB})=
 \frac{{\rm Tr}({\sf A})\,{\rm Tr}({\sf B})+{\bf A}\cdot{\bf B}}{2}
\end{equation}
for any two matrices ${\sf A}$ and ${\sf B}$.

As usual, we denote with ${\bf P}_{\bf p}$ and $\bar{\bf P}_{\bf p}$
the polarization vectors representing $\varrho_{\bf p}$ and
$\bar\varrho_{\bf p}$, respectively, and analogous expressions for
$\varrho_u\to{\bf P}_u$ and $\varrho_n\to{\bf P}_n$. The particle
and lepton-number matrices are represented by ${\sf S}\to {\bf S}$
and ${\sf D}\to {\bf D}$, in each case with the subscripts ${\bf
p}$, $u$ or $n$ for the different variables. Finally, we represent
the matrix of oscillation frequencies in the form
\begin{equation}
 \Omega=\frac{{\rm Tr}(\Omega)+\omega{\bf B}\cdot{\bm\sigma}}{2}\,,
\end{equation}
where
\begin{equation}
 {\rm Tr}(\Omega)=\frac{m_1^2+m_2^2}{2E}
 \hbox{\quad and\quad}
 \omega=\left|\frac{m_1^2-m_2^2}{2E}\right|\,.
\end{equation}
We have defined the oscillation frequency $\omega$ as a positive
number. The ``magnetic field'' is a unit vector that in the
interaction basis is explicitly ${\bf
B}=(\sin2\Theta,0,\cos2\Theta)$ where $\Theta$ is the vacuum mixing
angle. We have arbitrarily chosen its $y$-component to vanish,
corresponding to a choice of overall phase of the mixing matrix. In
the following, when we say that ``the mixing angle is small,'' we
mean that $|\sin2\Theta|\ll 1$. The sign of $B_x$ is physically
irrelevant whereas $B_z<0$ represents the normal, $B_z>0$ the
inverted hierarchy.

The equations of motion for the particle- and lepton-number
polarization vectors are explicitly
\begin{widetext}
\begin{eqnarray}\label{eq:eom6}
 \dot{\bf S}_n&=&\omega{\bf B}\times{\bf D}_n
 +\mu{\bf D}_0\times{\bf S}_n
 -\frac{\mu}{2}\,{\bf D}_1
 \times\left(a_n{\bf S}_{n-1}+b_n{\bf S}_{n+1}\right)\,,
 \nonumber\\
 \dot{\bf D}_n&=&\omega{\bf B}\times{\bf S}_n
 +\mu{\bf D}_0\times{\bf D}_n
 -\frac{\mu}{2}\,{\bf D}_1
 \times\left(a_n{\bf D}_{n-1}+b_n{\bf D}_{n+1}\right)\,.
\end{eqnarray}
\end{widetext}
The two lowest-order equations are explicitly
\begin{eqnarray}\label{eq:eom6a}
 \dot{\bf S}_0&=&\omega{\bf B}\times{\bf D}_0
 +\mu{\bf D}_0\times{\bf S}_0
 -\mu\,{\bf D}_1\times{\bf S}_{1}\,,
 \nonumber\\
 \dot{\bf D}_0&=&\omega{\bf B}\times{\bf S}_0
\end{eqnarray}
and
\begin{eqnarray}\label{eq:eom6b}
 \dot{\bf S}_1&=&\omega{\bf B}\times{\bf D}_1
 +\mu{\bf D}_0\times{\bf S}_1
 +{\textstyle\frac{1}{3}}\,\mu
 \left({\bf S}_{0}+2{\bf S}_{2}\right)\times{\bf D}_1\,,
 \nonumber\\
 \dot{\bf D}_1&=&\omega{\bf B}\times{\bf S}_1
 +{\textstyle\frac{2}{3}}\,
 \mu\left({\bf D}_{0}-{\bf D}_{2}\right)
 \times{\bf D}_1\,.
\end{eqnarray}

\clearpage

The conserved energy Eq.~(\ref{eq:energy2}) becomes, with the help
of Eq.~(\ref{eq:doubletrace}),
\begin{eqnarray}\label{eq:energy3}
 E&=&\frac{1}{2}\Bigl\lbrace{\rm Tr}(\Omega)\,{\rm Tr}({\sf S}_0)
 +\omega{\bf B}\cdot{\bf S}_0
 \nonumber\\
 &&{}+\frac{\mu}{2}\,
 \left[{\rm Tr}({\sf D}_0)^2-{\rm Tr}\,({\sf D}_1)^2
 +{\bf D}_0^2-{\bf D}_1^2\right]\Bigr\rbrace\,.
\end{eqnarray}
We have already noted that the individual traces ${\rm Tr}({\sf
S}_0)$, ${\rm Tr}({\sf D}_0)$ and ${\rm Tr}({\sf D}_1)$ are
conserved so that\footnote{The overall factor $1/2$ is missing in
the equivalent expression in our previous
paper~\cite{Hannestad:2006nj}.}
\begin{equation}\label{eq:energy4}
 E=\frac{1}{2}\left[\omega{\bf B}\cdot{\bf S}_0
 +\frac{\mu}{2}\,
 \left({\bf D}_0^2-{\bf D}_1^2\right)\right]
\end{equation}
is the conserved energy per neutrino, up to an irrelevant constant.
We use the notation
\begin{equation}
 E_{\omega}=\frac{\omega}{2}\left(|{\bf S}_0|+
 {\bf B}\cdot{\bf S}_0\right)
\end{equation}
for what is the ``potential energy'' in the pendulum analogy. Its
minimum is chosen at zero. The ``kinetic energy'' (neutrino-neutrino
interaction energy) is denoted by
\begin{equation}
 E_{\mu}=E_0+E_1=
 \frac{\mu}{4}\left({\bf D}_0^2-{\bf D}_1^2\right)\,,
\end{equation}
where we have introduced
\begin{eqnarray}
 E_0&=&+\frac{\mu}{4}\,{\bf D}_0^2\,,
 \nonumber\\
 E_1&=&-\frac{\mu}{4}\,{\bf D}_1^2
\end{eqnarray}
for the monopole and dipole contributions.

\section{Role of Anisotropy}                    \label{sec:anisotropy}

\subsection{Isotropic case}

\begin{figure*}
\begin{center}
\epsfig{file=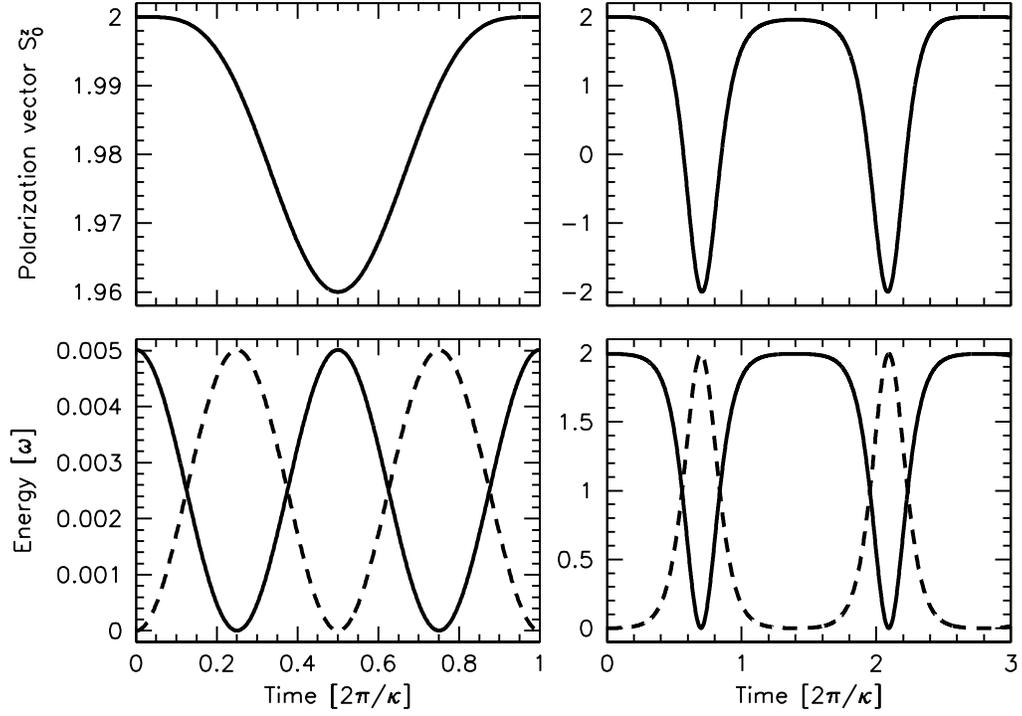,width=0.75\textwidth}
\end{center}
\caption{Evolution of a homogeneous and isotropic ensemble of equal
numbers of neutrinos and antineutrinos, initially in the electron
flavor, with a vacuum mixing angle $\sin2\Theta=0.1$. {\it Top:\/}
Polarization vector $S_0^z$. {\it Bottom:\/} Energy components
$E_\omega$~(solid) and $E_\mu$ (dashed). {\it Left:\/} Normal
hierarchy. {\it Right:\/} Inverted hierarchy.} \label{fig:isotropic}
\end{figure*}

In order to study the impact of kinematical decoherence in the
simplest nontrivial case we study a homogeneous ensemble consisting
of equal numbers of neutrinos and antineutrinos that are all
initially in the electron flavor. Before turning to issues of
decoherence, we briefly recall the evolution of the isotropic case.
To this end we show in Fig.~\ref{fig:isotropic} in the upper panels
the $z$-component of the overall polarization vector ${\bf S}_0={\bf
P}_0+\bar{\bf P}_0$ for the inverted and normal hierarchy. The
$\nu_e$ or $\bar\nu_e$ survival probability~is
\begin{equation}
 {\rm prob}({\nu_e\to\nu_e})=
 {\rm prob}(\bar\nu_e\to\bar\nu_e)=\frac{S_0^z+2}{4}\,,
\end{equation}
because in our normalization $|{\bf S}_0|=2$. In the lower panels we
show the corresponding energy components.

\begin{figure*}
\begin{center}
\epsfig{file=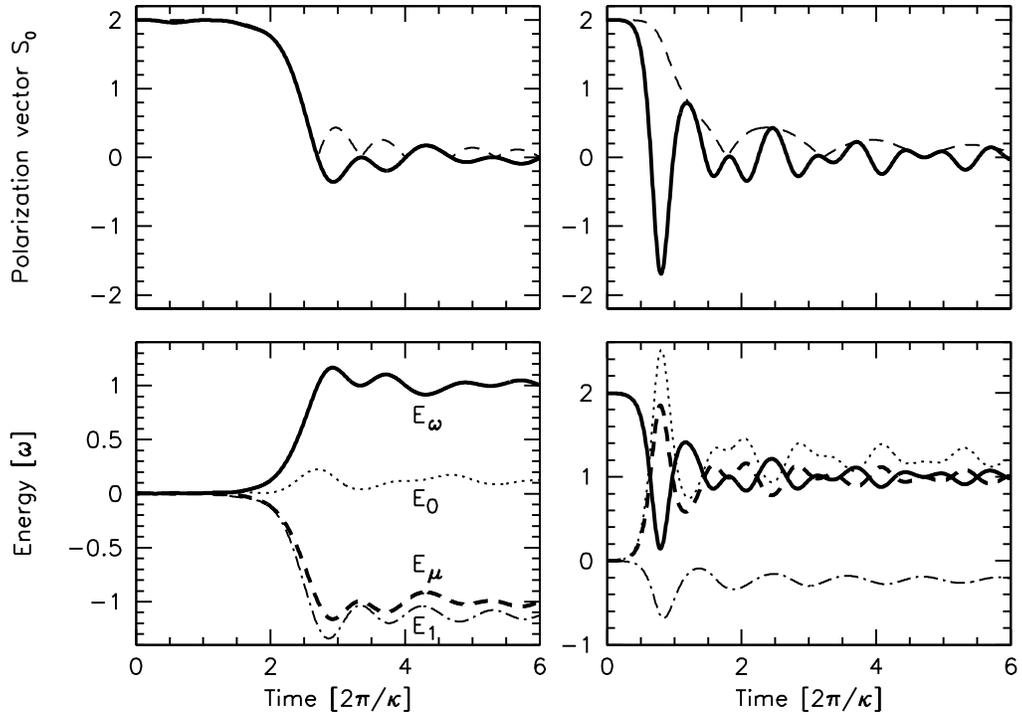,width=0.75\textwidth}
\end{center}
\caption{Same as Fig.~\ref{fig:isotropic}, now for a half-isotropic
neutrino gas. {\it Top:\/} Polarization vector $S_0^z$ (solid) and
the length $|{\bf S}_0|$ (dashed). {\it Bottom:\/} Energy components
$E_\omega$~(solid), $E_\mu=E_0+E_1$ (dashed), $E_0$ (thin dotted)
and $E_1$ (thin dot-dashed).} \label{fig:halfisotropic}
\end{figure*}

For the normal hierarchy (left) the motion is a small-excursion
harmonic oscillation with the bipolar frequency~$\kappa$. We show
one full cycle of oscillation. The potential energy $E_\omega$
(solid line in the lower panel) is measured against the mass
direction so that it is back at its maximum after a half-period as
behooves an ordinary pendulum. The dashed line shows the ``kinetic
pendulum energy'' $E_\mu$. The isotropy of the ensemble implies
$E_1=0$ and thus $E_\mu=E_0$.

For the inverted hierarchy (right) the motion is that of an inverted
oscillator. We show approximately one full cycle that for the chosen
mixing angle lasts approximately 3 times the bipolar period
$2\pi/\kappa$. The polarization vector starts in the positive $z$
direction, swings almost one full circle and almost arrives back at
the vertical direction (maximum in the middle of the plot), then
swings back, reaching its original vertical position, and then
starts over again. The maximum potential energy is $2\omega$ because
the energy is normalized to the total energy per $\nu$ and because
there are equal numbers of neutrinos and antineutrinos, i.e., two
particles per~$\nu$.

Reducing the mixing angle has the effect of reducing the oscillation
amplitude for the normal hierarchy, whereas it increases
logarithmically the duration of the ``plateau phases'' for the
inverted hierarchy. For quantitative discussions of the isotropic
case in terms of the pendulum analogy see
Ref.~\cite{Hannestad:2006nj}.

\subsection{Half-isotropic case}

Next we consider the same example, but assume a large degree of
anisotropy where only one half-space of momentum modes is occupied.
In Fig.~\ref{fig:halfisotropic} we show, in the upper panels, the
evolution of $S_0^z$. Moreover, we show the length of the overall
polarization vector, $|{\bf S_0}|$, as a thin dashed line. Both in
the normal hierarchy (left) and the inverted hierarchy (right), an
equal mixture of the two flavors is quickly achieved. In the
isotropic case the length of ${\bf P}_0$ and $\bar{\bf P}_0$ is
conserved and that of ${\bf S}_0$ is approximately conserved, up to
corrections of order $\omega/\mu\ll1$. Here, the lengths of ${\bf
P}_0$, $\bar{\bf P}_0$ and ${\bf S}_0$ shrink to zero, reflecting
kinematical decoherence. Of course, $S_0^z$ can become small or zero
without its length shrinking as during the first swing in the
inverted hierarchy case. It is the length of the polarization
vector, not its $z$-component, that is a measure of decoherence. Of
course, in the normal hierarchy, ${\bf S}_0$ performs only
small-excursion oscillations so that a significant change of its
$z$-component can be achieved only by a change of its length. In
this case $|{\bf S}_0|$ and $ S_0^z$ are almost identical.

The evolution of ${\bf S}_0$ was also illustrated in
Fig.~\ref{fig:trajectory} where we showed its trajectory in the
$x$-$z$-plane. For our chosen geometry where ${\bf B}$ is in the
$x$-$z$-plane, $S_0^y=0$ at all times. For the inverted hierarchy
(bottom) we indicate the isotropic-gas trajectory with diamonds at
time intervals of $0.01\,(2\pi/\kappa)$ where the motion starts in
the vertical position. In the half-isotropic case, the
particle-number polarization vector ${\bf S}_0$ spirals in to a
position close to the origin. The final offset is very small and
depends on the magnitude of $\omega/\mu$ for which we have chosen
$10^{-5}$. In the normal hierarchy (top), the isotropic-case motion
is a small-excursion harmonic oscillation between the dotted lines.
In the half-isotropic case, the evolution begins from the vertical
position. A full oscillation back to the vertical position is
performed before the shrinking of ${\bf S}_0$ becomes noticeable.

In the lower panels of Fig.~\ref{fig:halfisotropic} we show the
evolution of the different energy components. The simpler case is
the inverted hierarchy (right) where the ``potential'' and
``kinetic'' energies begin to oscillate as in the isotropic case of
Fig.~\ref{fig:isotropic}. On the time scale of a few bipolar
oscillation periods, the two components essentially equipartition,
although asymptotically a small offset remains that depends on
$\omega/\mu$. Moreover, the neutrino-neutrino energy (``kinetic
energy'') now develops a nonvanishing flux term $E_1=-\mu {\bf
D}_1^2/4$ that inevitably is negative.

The normal hierarchy (left) is initially similar in that $E_\omega$
and $E_\mu$ oscillate as for a pendulum, even though this motion is
not visible on the scale of the plot. As decoherence sets in, a
qualitatively different mode of behavior obtains in that the
neutrino-neutrino energy $E_\mu$ is dominated almost entirely by the
negative $E_1$ whereas $E_0$ now is subdominant. All individual
polarization vectors of all modes start aligned with the
$z$-direction, i.e., almost aligned with the force direction since
the mixing angle is small. Therefore, the initial $E_\omega$ is near
its minimum. If the overall polarization vector ${\bf S}_0$ is
supposed to shrink, the individual polarization vectors must develop
significant deviations from each other and thus the potential energy
must increase. Energy conservation then dictates that $E_\mu$
becomes negative.

One important conclusion is that the angular dependence of the
neutrino-neutrino interaction alone is not enough to cause
kinematical decoherence, but its absolute sign is also crucial. If
$E_1$ were not negative, energy conservation would prevent
significant decoherence for the normal hierarchy. Changing this sign
in a numerical example indeed reveals the absence of decoherence for
the normal hierarchy, but no significant change of behavior for the
inverted case. In the real world there is no choice about this sign.
It derives from the negative sign of the spatial part in the
neutrino current-current interaction, i.e., it is the negative sign
inherent in the Lorentz metric. From the equations of motion in the
form Eq.~(\ref{eq:eom6}) one would have never guessed that the
absolute sign of the term proportional to ${\bf D}_1$ plays a
crucial role.

\subsection{Evolution of multipoles}

It is also instructive to study the evolution of the higher
multipole components ${\bf S}_n$. We consider the same example as in
the previous section. For a half-isotropic gas the initial relative
length of the multipole components is
\begin{eqnarray}\label{eq:halfisotropiccoefficients}
 \frac{S_1}{S_0}&=&\frac{1}{2}\,,\nonumber\\
 \frac{S_{2m}}{S_0}&=&0\,,\nonumber\\
 \frac{S_{2m+1}}{S_0}&=&\frac{1}{2(m+1)}\,
 \prod_{k=1}^m\left(\frac{1}{2k}-1\right)
 \nonumber\\
 &=&\frac{(-1)^m}{2(m+1)}\,
 \frac{\Gamma(m+\frac{1}{2})}{\Gamma(\frac{1}{2})\Gamma(m+1)}\,,
\end{eqnarray}
where $S_n=|{\bf S}_n|$ and all of them are initially oriented in
the $z$-direction. Here, $m$ is an integer, i.e., the even
multipoles vanish initially except for $S_0$.

\begin{figure}[ht]
\begin{center}
\epsfig{file=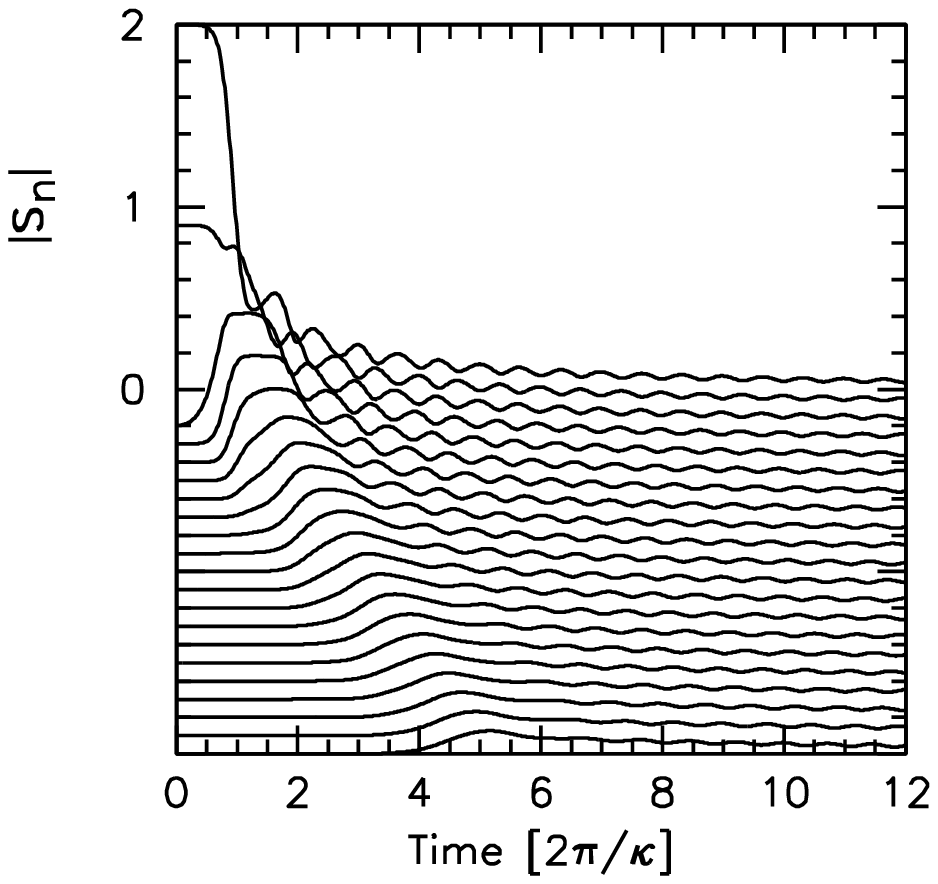,width=0.8\columnwidth}
\end{center}
\caption{Evolution of $|{\bf S}_n|$ for the first 21 multipoles in
the quasi-halfisotropic example with inverted hierarchy. The curves
are offset from each other by 0.1 vertical units.}
\label{fig:multipoleevolution}
\medskip
\begin{center}
\epsfig{file=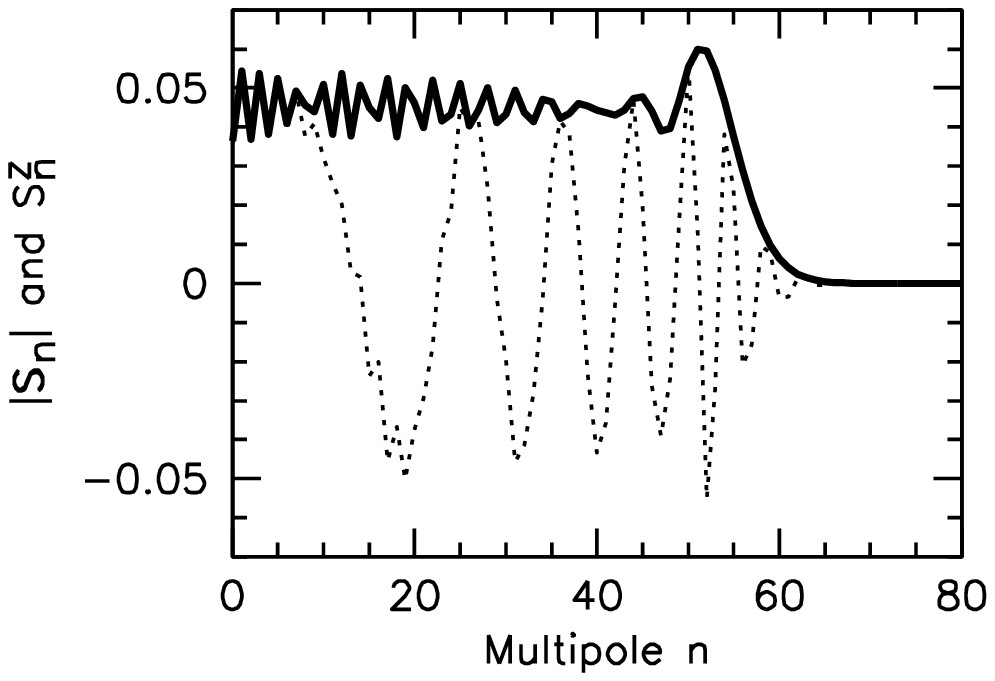,width=0.8\columnwidth}
\end{center}
\caption{$|{\bf S}_n|$ (solid) and $S_n^z$ (dotted) as a function of
$n$ at $t=12$ for the example of Fig.~\ref{fig:multipoleevolution}.}
\label{fig:multipoleevolution2}
\end{figure}

Instead of using these initial values, we calculate the evolution
for a case where all multipoles vanish initially except for $S_0=2$
and $S_1=1$, i.e., we use the same density and flux terms as in the
half-isotropic case. The evolution is very similar. We show the
evolution of the first 21 multipoles in
Fig.~\ref{fig:multipoleevolution} for the inverted hierarchy; the
picture is similar for the normal hierarchy. The initial horizontal
part of the curves corresponds to their zero initial value so that
the offset by 0.1 vertical units between the curves is directly
apparent.

Initially, $S_0$ and $S_1$ are large, but quickly decay away by
decoherence, whereas the higher multipoles vanish initially and get
excited one after another, but then decrease again. The length of
each multipole sports ``wiggles,'' i.e., the evolution is not
monotonic. Moreover, they have large relative angles in the $x$-$z$
plane (not shown here), i.e., the spread of the inital excitation to
larger and larger multipoles is far from simple in detail, but the
overall process is as expected. In
Fig.~\ref{fig:multipoleevolution2} we show $|{\bf S}_n|$ and $S_n^z$
as a function of $n$ at $t=12$ for the example of
Fig.~\ref{fig:multipoleevolution}. At $t=12$, multipoles $n\agt60$
are not yet excited. The nearly linear increase of the phases
between individual polarization vectors corresponds here to a
diffusion of the ``multipole wave'' to larger $n$ with approximately
constant speed, $n\simeq\kappa t$.

Qualitatively, this can be understood as follows. The second line in
Eq.~(\ref{eq:eom6}) together with the analogue of
Eq.~(\ref{eq:diffusion}) implies that
\begin{eqnarray}\label{eq:n_growth}
  \dot{\bf D}_n&\simeq&\omega{\bf B}\times{\bf S}_n
  -\mu{\bf D}_1\times\left(\frac{{\bf D}_n^\prime}{2n+1}
  +\frac{{\bf D}_n^{\prime\prime}}{2}\right)
  \nonumber\\
  &&\kern4em{}
  +\mu({\bf D}_0-{\bf D}_1)\times{\bf D}_n\,.
\end{eqnarray}
The last term does not contribute to the growth of $|{\bf D}_n|$ and
can be ignored. Approximate equipartition between potential and
kinetic energies in Eq.~(\ref{eq:energy4}) implies $|{\bf D}_0|
\sim|{\bf D}_1|\sim2(\omega/\mu)^{1/2}$, as indeed was observed in
Fig.~\ref{fig:halfisotropic}. The second term in
Eq.~(\ref{eq:n_growth}) then represents a combination of a drift in
$n-$space with velocity $\sim\kappa/(\sqrt2\,n)$ and diffusion with
diffusion coefficient $\sim\kappa/\sqrt2$. Up to a factor $t^{-1/2}$
this gives rise to an exponential factor
$\sim\exp[-n^2/(2\sqrt2\kappa t)]$. On the other hand, since $|{\bf
S}_n|\sim(\mu/\omega)^{1/2}|{\bf D}_n|$ for $n\geq2$, the first term
in Eq.~(\ref{eq:n_growth}) can give rise to exponential growth
whenever $({\bf B}\times{\bf S}_n)\cdot{\bf D}_n>0$. This growth
will be non-monotonic, but can roughly be estimated as
$\propto\exp(\kappa t/2)$ before ${\bf D}_n$ saturates. Using
perturbation theory, we will find a similar growth rate for ${\bf
D}_1$ at early times in Sect.~\ref{sec:analytic}. Combining these
two factors, one sees that at time $t$ the wave front of ${\bf
D}_n$, and thus also of ${\bf S}_n$, should be located at
$n\simeq\kappa t$, which, according to
Figs.~\ref{fig:multipoleevolution} and~\ref{fig:multipoleevolution2}
indeed it is within about $20$\%. The slow decrease of ${\bf S}_n$
and ${\bf D}_n$ at late times can be interpreted as due to the
$t^{-1/2}$ factor in the diffusive behavior.

\subsection{A simple model of decoherence}

The overall behavior of the inverted-hierarchy case is qualitatively
easy to understand. To this end we consider an ensemble of
polarization vectors ${\bf P}_u$, initially oriented in the
$z$-direction, that rotate around the $y$-axis, each with a
frequency $u$, so that
\begin{eqnarray}
 P^z_u&=&\cos(u t)\,,
 \nonumber\\
 P^x_u&=&\sin(u t)\,.
\end{eqnarray}
Both $u$ and $t$ are normalized to be dimensionless.

\begin{figure}
\begin{center}
\epsfig{file=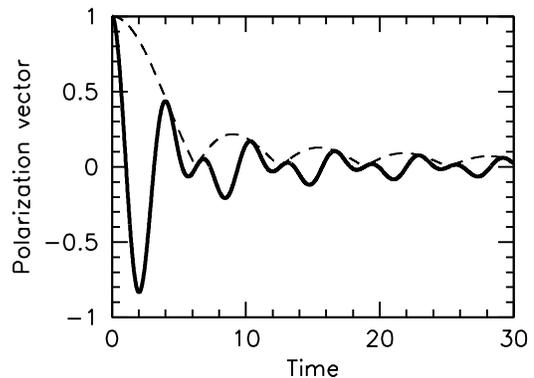,width=0.8\columnwidth}
\end{center}
\caption{Evolution of $P_0^z$ (solid) and $|{\bf P}_0|$ (dashed) of
our toy model with $\Delta u=1$.}\label{fig:toy1}
\end{figure}

\begin{figure}
\begin{center}
\epsfig{file=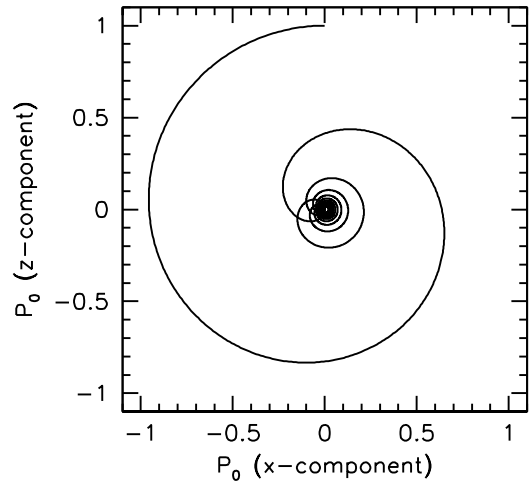,width=0.8\columnwidth}
\end{center}
\caption{Trajectory of ${\bf P}_0$ in the $x$-$z$-plane for our toy
model of decoherence with $\Delta u=1$.}\label{fig:toy2}
\end{figure}

If the frequencies are spread over $1\leq u\leq 1+\Delta u$, then
the overall polarization vector evolves as
\begin{eqnarray}
 P^z_0&=&\int_1^{1+\Delta u}\!\!du\,\cos(u t)
 =\frac{\sin[(1+\Delta u)\,t]-\sin(t)}{t}\,,
 \nonumber\\
 P^x_0&=&\int_1^{1+\Delta u}\!\!du\,\sin(u t)
 =\frac{\cos(t)-\cos[(1+\Delta u)\,t]}{t}\,,
 \nonumber\\
 |{\bf P}_0|&=&\left[(P^x_0)^2+(P^z_0)^2\right]^{1/2}
 = \frac{2\,|\sin(\Delta u\,t/2)|}{t}\,.
\end{eqnarray}
For $\Delta u=1$ we show in Fig.~\ref{fig:toy1} the evolution of
$P_0^z$ (solid line) and the length $|{\bf P}_0|$ (dashed line). In
Fig.~\ref{fig:toy2} we show the trajectory of ${\bf P}_0$ in the
$x$-$z$-plane.


Unsurprisingly, this simple model represents qualitatively the
features of the inverted-hierarchy case. In particular, the
trajectory in the $x$-$z$-plane is not a true spiral, but there are
many crossings of the origin as the length of the overall
polarization vector shrinks. Its length does not shrink
monotonically. Its envelope decreases as a power law $t^{-1}$, not
exponentially.

This example also shows that it is not trivial to define a useful
measure of kinematical decoherence. The $z$-component of the
polarization vector is not useful because ${\bf P}_0$ can tilt so
that its $z$-component vanishes, i.e., both flavors are present with
the same probability, yet the flavor content of the ensemble is
perfectly coherent. The length of ${\bf P}_0$ is a much better
measure in analogy to dynamical decoherence of a single mode where
the length of ${\bf P}_{\bf p}$ is a measure of dynamical
decoherence; unit length would correspond to a pure state. In our
case, the length of the individual ${\bf P}_{\bf p}$ is conserved,
whereas the length of the total ${\bf P}_0$ shrinks, but not
monotonically.

It is not evident if there is an ``entropy'' measure that evolves
monotonically as a consequence of kinematical decoherence. We note,
however, that the origin of kinematical decoherence is the
differential evolution between neighboring polarization vectors. In
our toy example, the angle between neighboring polarization vectors
grows linearly as $\Delta u\,t$ and thus is a measure of
decoherence.

\begin{figure}[ht]
\begin{center}
\epsfig{file=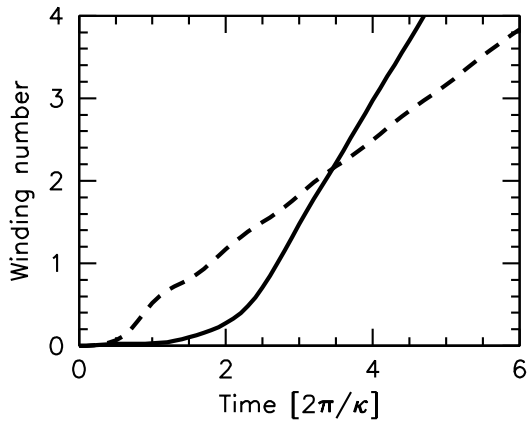,width=0.8\columnwidth}
\end{center}
\caption{Evolution of the ``winding number'' as defined in
  Eq.~(\ref{eq:windingnumber}) for the half-isotropic case of
  Fig.~\ref{fig:halfisotropic}. Normal hierarchy (solid) and inverted
  hierarchy (dashed).}\label{fig:windingnumber}
\end{figure}

Motivated by this observation we can define the ``winding
number''
\begin{equation}\label{eq:windingnumber}
 N=\frac{1}{2\pi}\int_{-1}^{+1}du\,
 \frac{|d{\bf P}_u/du|}{|{\bf P}_u|}\,.
\end{equation}
In the symmetric case where the evolution of the polarization
vectors is essentially restricted to the $x$-$z$-plane, this
quantity tells us the number of windings around the $y$-direction of
the full ensemble of polarization vectors. In
Fig.~\ref{fig:windingnumber} we show the evolution of this quantity
for the normal and inverted hierarchy examples of
Fig.~\ref{fig:halfisotropic}. $N$~counts how often the length of
${\bf S}_0$ shrinks to zero.

\subsection{Recurrence effects}

Closely related to kinematical decoherence are recurrence effects
that arise when one uses a limited number of polarization vectors,
i.e., when the full ensemble is represented by a coarse-grained
(binned) ensemble of polarization vectors or, in multipole space,
when the series is truncated at some multipole $n_{\rm max}$.
Binning or truncation are unavoidable in numerical treatments
and represent equivalent approximations.\footnote{In
Ref.~\cite{Sawyer:2005jk} a multipole expansion (``partial wave
expansion'') was proposed and dismissed with the argument that the
necessary truncation would introduce uncontrolled numerical effects
whereas coarse bins in angle space would provide limited resolution,
yet a correct solution of the nonlinear equations of motion. We
disagree with this assessment. The same unphysical recurrence
effects occur both if the number of angular bins or the multipole
order of truncation is too small.}

\begin{figure}[b]
\begin{center}
\epsfig{file=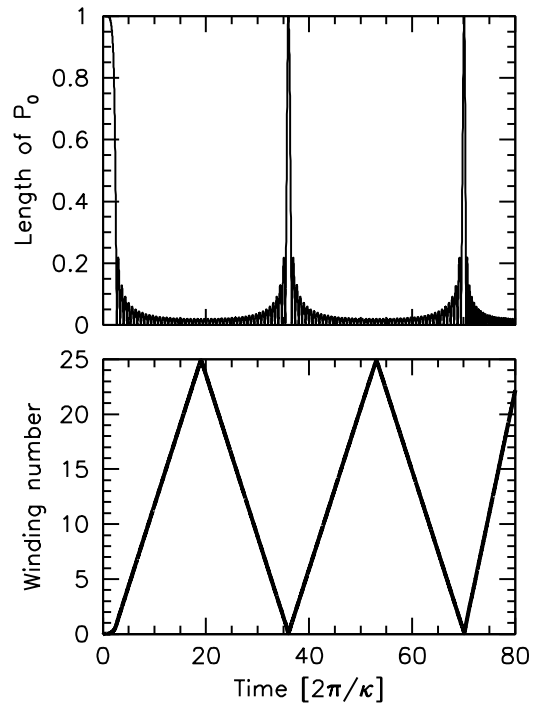,width=0.8\columnwidth}
\end{center}
\caption{Evolution of
  $|{\bf P}_0|$ and the winding number for the normal-hierarchy case
  of Fig.~\ref{fig:halfisotropic} with 51 polarization
  vectors.}\label{fig:recurrence}
\end{figure}

If the phase differences evolve linearly between different
polarization vectors as in the toy example of the previous section,
then the initial state will recur when each polarization vector has
turned around the $y$-axis by $2\pi$ relative to its neighbors. Put
another way, if we use $n_{\rm pol}$ bins to represent the
polarization vectors, we expect recurrence to begin when $N=(n_{\rm
pol}-1)/2$. We demonstrate this effect in Fig.~\ref{fig:recurrence}
for the same example as in Fig.~\ref{fig:halfisotropic}, normal
hierarchy, using $n_{\rm pol}=51$. Indeed, the winding number
increases until approximately $N=25$ and then decreases back to
almost zero, and so forth. This system is almost periodic on the
recurrence time scale, reflecting that nonlinear effects play a
subleading role here. Since a large flux term is present from the
start, kinematical decoherence is very similar to the linear toy
example of the previous section where recurrence would be exact and
the system would be periodic with a frequency proportional to
$(n_{\rm pol}-1)^{-1}$ times the individual polarization vector's
oscillation frequency.

Using the multipoles as the primary variables, the same
recurrence effects occur due to the truncation $n_{\rm max}$ of the
multipole series. The ``multipole wave'' of
Fig.~\ref{fig:multipoleevolution2} can only propagate up to $n_{\rm
max}$, is then reflected, and eventually returns back to $n=0$. This
is the moment of maximum recurrence of the overall polarization
vector.

For a more complicated example such as a realistic simulation of the
supernova case, recurrence effects may not be so obvious as here.
Therefore, in a numerical treatment one must make sure that the
largest relative angle developed between any two neighboring
polarization vectors never grows to order unity, i.e., one should
monitor the largest relative angle between neighboring polarization
vectors as a measure of accuracy of the calculation. In other words,
kinematical decoherence is a property of the entire ensemble, but
can also be a a property of individual bins if they are too coarse.
One must make sure that within all individual bins kinematical
decoherence remains negligible.

\section{Small Initial Anisotropy}         \label{sec:smallanisotropy}

\subsection{Numerical example}

We next turn to a symmetric system that is perfectly isotropic
except for an arbitrarily small but nonzero initial anisotropy. We
work in the multipole picture and assume that initially ${\bf
D}_n=0$ for all $n$, ${\bf S}_n=0$ for $n\geq 2$, $S_0^z=2$ and
$S_1^z=\xi\ll1$. Note that $\xi=1$ corresponds to a flux term
equivalent to the half-isotropic case.

\begin{figure*}
\begin{center}
\epsfig{file=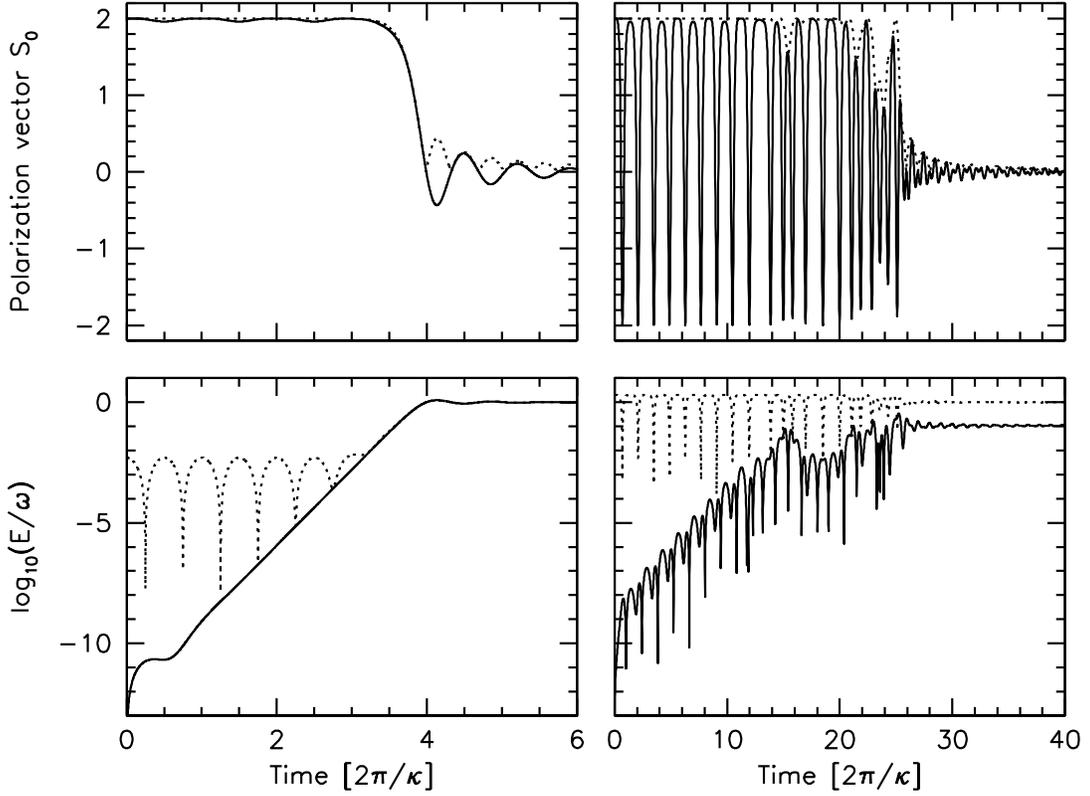,width=0.8\textwidth}
\end{center}
\caption{Evolution of a homogeneous and near-isotropic ensemble of
equal $\nu$ and $\bar\nu$ densities with $\sin2\Theta=0.1$ as in
Figs.~\ref{fig:isotropic} and~\ref{fig:halfisotropic}. The initial
anisotropy is $\xi=10^{-4}$. {\it Top:\/} Polarization vector
$S_0^z$ (solid) and the length $|{\bf S}_0|$ (dotted). {\it
Bottom:\/} Neutrino-neutrino flux energy $-E_1$ (solid) and
potential energy $E_\omega$~(dotted). {\it Left:\/} Normal
hierarchy. {\it Right:\/} Inverted hierarchy.}
\label{fig:nearisotropic}
\end{figure*}

As a first example we show in Fig.~\ref{fig:nearisotropic} the
evolution of the overall polarization vector ${\bf S}_0$ (top) for
the normal hierarchy (left) and the inverted hierarchy (right). The
initial anisotropy is $\xi=10^{-4}$ and $\sin2\Theta=0.1$. For both
hierarchies, kinematical decoherence eventually obtains.

In the bottom panels we show the evolution of $-E_1=(\mu/4)\,{\bf
D}_1^2$ (solid line) as a proxy for ${\bf D}_1$ which is the primary
quantity that causes decoherence. For the symmetric neutrino gas
only the component $D_1^y$ is nonzero, but it can be negative and
can change sign so that it is difficult to display on a logarithmic
scale. We also show the evolution of the ``potential energy''
$E_\omega$ (dotted line).

For the normal hierarchy, $D_1^y$ grows exponentially after a short
initial transient. While $D_1^y$ is small, the potential energy
$E_\omega$ performs the usual oscillations of a harmonic oscillator,
complemented by the opposite oscillations of the kinetic energy
$E_\mu=E_0-E_1$ (not shown) that is entirely dominated by $E_0$.  As
$D_1^y$ grows, it eventually dominates the kinetic energy and the
end state is approximately $E_\omega=+\omega$ and $E_\mu=-\omega$,
with $E_1$ dominating $E_\mu$. In other words, the end state is
exactly as in the previous case with a large initial anisotropy. The
crucial novel feature is that the flux term grows exponentially from
a small value. This is a purely nonlinear effect caused by
neutrino-neutrino interactions. For smaller mixing angles and/or
other values of $\xi$, the behavior is analogous. The rising part of
the $E_1$ curve is the same, except that it shifts vertically in
direct proportion to $\xi$ and $\sin2\Theta$.

For the inverted hierarchy, the evolution is far more complicated.
While $D_1^y$ grows in an average sense, this evolution is overlaid
with oscillations and in fact $D_1^y$ changes sign at each spike in
the $E_1$ curve. Moreover, the evolution of the envelope of the
$E_1$ curve is not monotonic. It can perform complicated motions,
with growing, declining, and nearly flat phases, until finally
decoherence obtains.

From these numerical observations we tentatively conclude that an
isotropic neutrino ensemble of equal densities of neutrinos and
antineutrinos is not stable. A small flux term triggers a run-away
evolution towards kinematical decoherence.

\subsection{Analytic treatment}\label{sec:analytic}

\subsubsection{Simplified equations of motion}

As a starting point for an analytic understanding of this
instability we use the equations of motion Eq.~(\ref{eq:eom6}) for
the vectors ${\bf S}_n$ and ${\bf D}_n$. Initially only ${\bf S}_0$
is of order one, $|{\bf S}_1|=\xi\ll1$, and all others vanish. As
long as ${\bf D}_1$ is sufficiently small, none of the higher
multipoles can grow large so that we can limit our attention to the
equations for $n=0$ and $n=1$, i.e., to Eqs.~(\ref{eq:eom6a})
and~(\ref{eq:eom6b}).

We further observe that the $n=0$ equations are coupled to the
higher multipoles only by the term $-\mu{\bf D}_1\times{\bf S}_1$,
consisting of a product of two small quantities, i.e., it is at
least of order $\xi^2$. Therefore, the evolution of the
near-isotropic ensemble is identical with that of the isotropic case
until ${\bf D}_1$ has grown sufficiently large. In other words, up
to second order in $\xi$ the $n=0$ equation is that of the isotropic
case.

Therefore, what remains to be solved is the $n=1$ equation. After
neglecting terms involving $n\geq 2$ all we need to study is
\begin{eqnarray}\label{eq:eom6c}
 \dot{\bf S}_1&=&
 \left(\omega{\bf B}+\frac{\mu}{3}\,{\bf S}_0\right)
 \times{\bf D}_1
 +\mu{\bf D}_0\times{\bf S}_1\,,
 \nonumber\\
 \dot{\bf D}_1&=&\omega{\bf B}\times{\bf S}_1
 +\frac{2\mu}{3}\,{\bf D}_{0}\times{\bf D}_1\,,
\end{eqnarray}
where ${\bf S}_0(t)$ and ${\bf D}_0(t)$ are the solutions of the
unperturbed pendulum equations.

The equations simplify further in our case of equal $\nu$ and
$\bar\nu$ densities where symmetry dictates that all polarization
vectors ${\bf P}_{\bf p}$ and $\bar{\bf P}_{\bf p}$ evolve as each
other's mirror images relative to the $x$-$z$-plane, the latter
being singled out by our choice that ${\bf B}$ lies in that plane.
In this case all ${\bf D}_n$ vectors are parallel to the $y$-axis,
whereas all ${\bf S}_n$ vectors are confined to the $x$-$z$-plane.

As a consequence, the ${\bf D}_0\times{\bf D}_1$ term drops out. The
second equation then implies $\ddot{\bf D}_1=\omega{\bf
B}\times\dot{\bf S}_1$, and with the first equation yields
\begin{equation}
 \ddot{\bf D}_1=\omega{\bf B}\times
 \left[\left(\omega{\bf B}+\frac{\mu}{3}\,{\bf S}_0\right)
 \times{\bf D}_1\right]
 +\omega\mu{\bf B}\times\left({\bf D}_0\times{\bf S}_1\right)\,.
\end{equation}
Expanding the triple product, observing that in our case ${\bf
B}\cdot{\bf D}_n=0$, and using ${\bf B}^2=1$ this is
\begin{equation}\label{eq:ddot}
 \ddot{\bf D}_1=-
 \left(\omega^2+\frac{\omega\mu}{3}\,{\bf B}\cdot{\bf S}_0\right)
 {\bf D}_1
 +\omega\mu({\bf B}\cdot{\bf S}_1){\bf D}_0\,.
\end{equation}
We repeat that all ${\bf D}_n$ are here parallel to each other and
to the $y$-axis. In the absence of neutrino-neutrino interactions
($\mu=0$), ${\bf D}_1$ obeys a harmonic-oscillator equation.

In a dense neutrino gas, $\mu\gg\omega$, we may ignore the term
proportional to $\omega^2$. Since during the growth phase, $|{\bf
D}_1|\lesssim|{\bf D}_0|$ and $|{\bf S}_1|\ll|{\bf S}_0|$, the last
term in Eq.~(\ref{eq:ddot}) can also be neglected. A possible
exponential growth is then accounted for by the term
\begin{equation}\label{eq:expgrowth1}
 \ddot{\bf D}_1=-\frac{\kappa^2}{3}\,
 \frac{{\bf B}\cdot{\bf S}_0}{2}\,{\bf D}_1\,,
\end{equation}
where we have used the bipolar oscillation frequency
$\kappa^2=2\omega\mu$ and we note that initially $|{\bf S}_0|=2$ and
that its length remains nearly constant if $|\sin2\Theta|\ll1$
unless there is kinematical decoherence.

\subsubsection{Normal hierarchy}

If the mixing angle is small, the ``normal hierarchy'' is defined by
neutrinos being essentially in the lower mass eigenstate or ${\bf
B}\cdot{\bf S}_0/2=-1$ up to corrections of order $\sin^22\Theta$.
Therefore, the critical part of the equation of motion for the flux
term is approximately
\begin{equation}\label{eq:expgrowth2}
 \ddot{\bf D}_1=+\frac{\kappa^2}{3}\,{\bf D}_1\,.
\end{equation}
The solution includes growing modes of the form
\begin{equation}\label{eq:expgrowth3}
 {\bf D}_1\propto \exp\left(+\frac{2\pi}{\sqrt3}\,\tau\right)\,,
\end{equation}
where $\tau=\kappa t/2\pi$ is the dimensionless time variable used
in all of our plots, i.e., time in units of the bipolar oscillation
period. This behavior is numerically verified in the lower-left
panel of Fig.~\ref{fig:nearisotropic} if we recall that the plotted
quantity is $-E_1\propto{\bf D}_1^2\propto
\exp[(4\pi/\sqrt3)\,\tau]$.

Once more we note that the crucial absolute sign in
Eq.~(\ref{eq:expgrowth2}) is traced back to the negative sign in the
term $(1-{\bf v}_{\bf q}\cdot{\bf v}_{\bf p})$ in
Eq.~(\ref{eq:eom1}). Therefore, the instability of the
near-isotropic neutrino gas, for the normal hierarchy, is not just
caused by different angular modes experiencing different
neutrino-neutrino effects. This would also be the case for a
hypothetical term $(1+{\bf v}_{\bf q}\cdot{\bf v}_{\bf p})$ that
naively looks very similar, but causes a completely different
overall behavior. In more complicated numerical situations, e.g.,
for a full supernova simulation, it is easy to change this sign in
order to diagnose the relevance of the term
Eq.~(\ref{eq:expgrowth1}) for the overall behavior of the system.

\begin{figure}[b]
\begin{center}
\epsfig{file=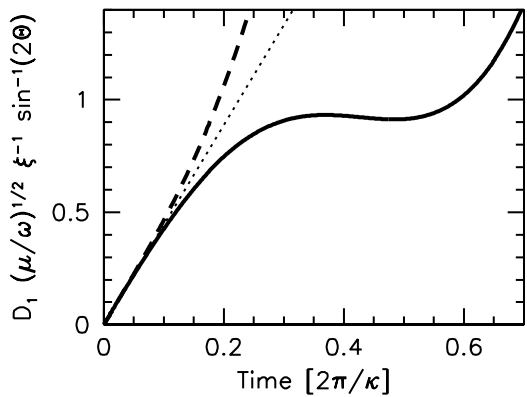,width=0.8\columnwidth}
\end{center}
\caption{Initial evolution of ${\bf D}_1$ as discussed in the text
for the normal hierarchy (solid) and the inverted hierarchy
(dashed). The dotted line is the linear relationship of
Eq.~(\ref{eq:initial2}).}\label{fig:initialevolution}
\end{figure}

Initially ${\bf D}_1=0$ so that the exponential growth term alone is
not enough. The initial evolution is dominated by the second line of
Eq.~(\ref{eq:eom6c}) that is initially
\begin{equation}\label{eq:initial1}
 \dot D_1 =\left(\frac{\omega}{\mu}\right)^{1/2}
 \frac{\kappa}{\sqrt2}\,\,\xi\sin2\Theta\,,
\end{equation}
where $D_1=|{\bf D}_1|$ and we have used $|{\bf B}|=1$ and ${\bf
S}_1(0)=(0,0,\xi)$. Here the absolute sign is not critical because
it does not matter if ${\bf D}_1$ grows in the positive or negative
$y$-direction. The first factor $(\omega/\mu)^{1/2}$ sets the
overall scale for ${\bf D}_1$. From the expression for the energy we
note that ${\bf S}$ and ${\bf D}$ appear in the combinations $\omega
|{\bf S}_0|$ and $\mu{\bf D}_0^2$ or $\mu{\bf D}_1^2$. The
quantities ${\bf S}_n$ are of order unity, whereas the natural scale
for ${\bf D}_n$ is $(\omega/\mu)^{1/2}$. The initial evolution is
\begin{equation}\label{eq:initial2}
 \left(\frac{\mu}{\omega}\right)^{1/2}
 \frac{D_1}{\xi\sin2\Theta}
 =\frac{2\pi}{\sqrt2}\,\tau\,,
\end{equation}
where, again, $\tau=(\kappa/2\pi)\,t$ is our usual measure of time.
This linear growth is shown in Fig.~\ref{fig:initialevolution} as a
dotted line and agrees with the numerical examples.

Besides a transient caused by the other terms in the equation, the
exponential growth takes over after a time of order $2\pi/\kappa$.
We conclude that the initial value of $(\mu/\omega)^{1/2}D_1$ for
its exponential growth is of order $\xi\sin(2\Theta)$ so that it
grows to order unity within a time scale of order
$-(2\pi/\kappa)\,\ln(\xi\sin2\Theta)$.

\subsubsection{Inverted hierarchy}

For the inverted hierarchy, the initial growth of $D_1$ is the same
(dashed line in Fig.~\ref{fig:initialevolution}). After that,
however, the situation is entirely different because initially ${\bf
B}\cdot{\bf S}_0/2=+1$. This corresponds to the pendulum starting in
a nearly upright position from where it starts almost full-circle
oscillations. For our usual assumption $|\sin2\Theta|\ll 1$ the term
${\bf B}\cdot{\bf S}_0$ is almost identical with $S_0^z$ shown in
the upper-right panel of Fig.~\ref{fig:nearisotropic}. This term
changes sign during every swing of the pendulum. During the phases
when ${\bf B}\cdot{\bf S}_0>0$, ${\bf D}_1$ oscillates whereas
during the phases ${\bf B}\cdot{\bf S}_0<0$ exponential growth
obtains. These are the relatively short phases when the pendulum is
oriented downward, whereas it stays upright for long periods if the
mixing angle is small.

\begin{figure}[ht]
\begin{center}
\epsfig{file=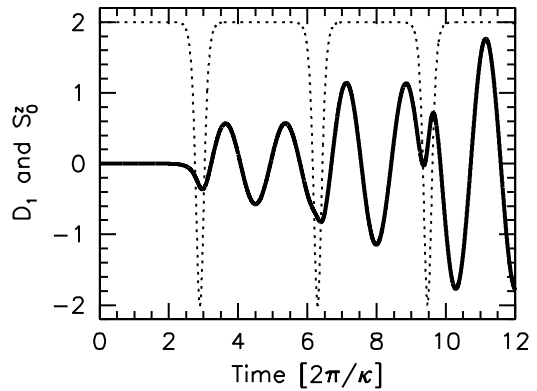,width=0.8\columnwidth}
\end{center}
\caption{Initial evolution of ${\bf D}_1$ (solid line, arbitrary
units) and $S_0^z$ (dotted) for the inverted hierarchy with
$\xi=10^{-4}$ and
$\sin2\Theta=10^{-7}$.}\label{fig:initialevolution2}
\end{figure}

We illustrate this behavior in Fig.~\ref{fig:initialevolution2}
where we show the evolution of $D_1$ overlaid with that of $S_0^z$.
We have chosen an extremely small mixing angle,
$\sin2\Theta=10^{-7}$, to obtain long plateaus for $S_0^z$ where the
pendulum stands almost still in an almost upright position. During
these phases we have ${\bf B}\cdot{\bf S_0}/2=+1$ so that
Eq.~(\ref{eq:expgrowth1}) implies harmonic oscillations of $D_1$
with the frequency $\kappa/\sqrt3$ as borne out by the example in
Fig.~\ref{fig:initialevolution2}.

The frequency $\kappa/\sqrt3$ is not matched to the pendulum's
oscillation period because the duration of the ``plateaus'' depend
logarithmically on $\sin2\Theta$. Therefore, the short exponential
growth phases when $S_0^z<0$ occur at erratic instances relative to
the harmonic ${\bf D}_1$ oscillation. This interplay explains the
erratic behavior of the ${\bf D}_1$ evolution that is apparent in
Fig.~\ref{fig:nearisotropic} and that obtains in all numerical
examples. This interplay also explains why small changes of
parameters such as $\sin2\Theta$ can completely change the overall
${\bf D}_1$ evolution.

Therefore, we cannot prove if there is some specific combination of
parameters where kinematical decoherence will not occur, although
this would have to be isolated parameter points that presumably have
measure zero in parameter space if they exist at all.

Except for this caveat we conclude that both for the normal and
inverted hierarchy kinematical decoherence is an unavoidable
consequence of the nonlinear neutrino-neutrino interaction terms. An
infinitesimally small, but nonzero, deviation from isotropy is
enough to trigger an exponential evolution towards flavor
equilibrium. Therefore, the pendulum solution that describes the
behavior of a perfectly isotropic gas is merely an unstable limit
cycle of this nonlinear system.

\section{Conclusions and outlook}
\label{sec:conclusions}

We have investigated multi-angle kinematical decoherence effects in
a dense neutrino gas consisting of equal $\nu$ and $\bar\nu$
densities. The current-current nature of the weak-interaction
Hamiltonian implies that a current of the background medium causes
kinematical decoherence between neutrinos propagating in different
directions. This simple effect becomes entirely nontrivial in the
most interesting case when the ``background current'' is caused by
the neutrinos themselves.

We have shown that a neutrino gas of this sort is not stable in
flavor space. If one prepares the ensemble in a given flavor state,
even a small deviation from isotropy is enough to trigger an
exponential evolution towards flavor equipartition, even when the
mixing angle is small. Up to logarithmic corrections, the time scale
is determined by the bipolar oscillation period. We thus have to do
with a self-induced macroscopic pair conversion process that
proceeds much faster than ordinary pair processes that occur at
order $G_{\rm F}^2$.

With hindsight this is probably the same ``speed-up effect'' discussed
in Ref.~\cite{Sawyer:2005jk}. However, the fast speed
$\kappa=\sqrt{2\omega\mu}$ is not introduced by the multi-angle effect
as is perhaps suggested in Ref.~\cite{Sawyer:2005jk}. Rather, $\kappa$
is the bipolar oscillation frequency that is inherent even in the
isotropic system. For the normal hierarchy ``nothing'' seems to happen
in the isotropic case, although this would be a misperception because
the system moves with the speed~$\kappa$, but the amplitude is small
if $\sin2\Theta\ll1$. In any event, we agree with the importance of
multi-angle effects and with the end result conjectured in
Ref.~\cite{Sawyer:2005jk} that complete flavor equipartition obtains
on a time scale~$\kappa^{-1}$.  However, we have found no indication
that an even faster time scale $\mu^{-1}$ plays any role, in contrast
to what was conjectured in Ref.~\cite{Sawyer:2005jk}.

Of course, flavor equipartition obtains only in a macroscopic sense
with some degree of coarse graining in phase space. It is the nature
of kinematical decoherence that information is not lost and the
entropy does not increase. On the other hand, kinematical and
dynamical decoherence are not always operationally distinguishable.
It is not necessarily possible to distinguish between a neutrino gas
in true chemical equilibrium and one where neighboring modes are
simply ``de-phased'' relative to each other.

Either way, the ``self-maintained coherence'' of a dense neutrino
gas represents only an unstable limiting form of behavior of the
perfectly isotropic case. Since no gas can be perfectly isotropic,
self-induced decoherence is the generic behavior of a symmetric
$\nu\bar\nu$ gas.

We have only analyzed the two-flavor case. It remains to be studied
how our results carry over to a genuine three-flavor situation.

Contrary to a naive expectation, the fact that neutrinos propagating
in different directions experience different weak potentials is
necessary, but not sufficient, to cause multi-angle decoherence. The
negative sign in the Lorentz metric plays a crucial role because it
determines that the flux term grows exponentially rather than
oscillating. The occurrence of kinematical decoherence is a subtle
feature of the nonlinear equations.

We have studied the simplest possible case where decoherence effects
obtain. A more realistic case is provided by neutrinos streaming off
a supernova core. In this case the fluxes of neutrinos and
antineutrinos are different and geometry implies that the
neutrino-neutrino interaction strength declines with the fourth
power of radius. The evolution of an asymmetric system is far more
complicated, even in the isotropic case, where the simple pendular
motion turns into one of a spinning top that can precess, nutate, or
simply swing, depending on the $\nu\bar\nu$ asymmetry. In the
supernova context, transitions between these modes occur as $\mu$
declines as a function of radius. Multi-angle decoherence cannot
occur, for example, in regions where synchronized oscillations
prevail.

Numerical simulations suggest that for neutrinos streaming off a
supernova core, the collective behavior characteristic of an
isotropic gas prevails in many situations, i.e., kinematical
decoherence occurs, but is not the dominating feature of the system,
contrary to what we have found here for the symmetric $\nu\bar\nu$
gas. Evidently it is not straightforward to apply the insights
gained from our present study to the supernova case.

We imagine, however, that the methods developed here may take us one
step further to understand the supernova problem. We have formulated
the equations of motion in a novel form adapted to the problem of
multi-angle propagation, i.e., we have used a multipole expansion
rather than the usual momentum modes and we have used the particle
number and lepton number polarization vectors (or density matrices)
as the fundamental variables. In this way, the crucial exponentially
growing quantity, our ${\bf D}_1$, could be isolated as the primary
cause of kinematical decoherence. This approach may prove useful for
understanding the supernova problem as well.

For numerical three-flavor studies we have provided an expression
for the conserved energy of the system in terms of the density
matrices. The conserved energy and its three components $E_\omega$,
$E_0$ and $E_1$ provide a useful diagnostic tool for the behavior of
the system.

Even simple nonlinear systems can exhibit a surprisingly rich
phenomenology. It is fascinating that the neutrinos streaming off a
supernova core provide an intriguing case in point where many
aspects of its behavior remain to be understood.



\begin{acknowledgments}
This work was partly supported by the Deutsche
Forschungsgemeinschaft under Grants No.~SFB-375 and TR-27 and by the
European Union under the ILIAS project, contract
No.~RII3-CT-2004-506222.
\end{acknowledgments}


\end{document}